\documentclass[aps,pra,preprint,superscriptaddress]{revtex4-1}

\usepackage{color}
\usepackage{graphicx}
\usepackage{amsmath}
\usepackage{amsfonts}

\begin{document}


\title{Off-axis vortex beam propagation through classical optical system in terms of Kummer confluent hypergeometric function}


\author{Ireneusz Augustyniak}
\affiliation{Department of Applied Mathematics, Faculty of Pure and Applied Mathematics, Wroclaw University of Science and Technology, Wybrze\.ze Wyspia\'nskiego 27, 50-370 Wroc\l{}aw, Poland}

\author{Weronika Lamperska}
\affiliation{Department of Optics and Photonics, Faculty of Fundamental Problems of Technology, Wroclaw University of Science and Technology, Wybrze\.ze Wyspia\'nskiego 27, 50-370 Wroc\l{}aw, Poland}

\author{Jan Masajada}
\affiliation{Department of Optics and Photonics, Faculty of Fundamental Problems of Technology, Wroclaw University of Science and Technology, Wybrze\.ze Wyspia\'nskiego 27, 50-370 Wroc\l{}aw, Poland}

\author{\L{}ukasz P\l{}ociniczak}
\affiliation{Department of Applied Mathematics, Faculty of Pure and Applied Mathematics, Wroclaw University of Science and Technology, Wybrze\.ze Wyspia\'nskiego 27, 50-370 Wroc\l{}aw, Poland}

\author{Agnieszka Popio\l{}ek-Masajada}
\affiliation{Department of Optics and Photonics, Faculty of Fundamental Problems of Technology, Wroclaw University of Science and Technology, Wybrze\.ze Wyspia\'nskiego 27, 50-370 Wroc\l{}aw, Poland}

\date{\today}

\begin{abstract}
The analytical solution for the propagation of the laser beam with optical vortex through the system of lenses is presented. The optical vortex is introduced into the laser beam (described as Gaussian beam) by spiral phase plate. The solution is general as it holds for the optical vortex of any integer topological charge, the off-axis position of the spiral phase plate and any number of lenses. Some intriguing conclusions are discussed. The higher order vortices are unstable and split under small phase or amplitude disturbance. Nevertheless, we have shown that off-axis higher order vortices are stable during the propagation through the set of lenses described in paraxial approximation, which is untypical behavior. The vortex trajectory registered at image plane due to spiral phase plate shift behaves like a rigid body. We have introduced a new factor which in our beam plays the same role as Gouy phase in pure Gaussian beam. 
\end{abstract}


\maketitle

\section{Introduction\label{sect:Introduction}}
The question of propagation of light fields containing optical vortices \cite{1:Vasnetsov,2:Soskin,3:Dennis} is getting attention in many fields of modern optical science nowadays. The fundamental case of such a field is vortex beam - the well-defined, single beam (as for example LG beam \cite{4:Kotlyar,5:Arora}), carrying the optical vortex of any order. Many different problems concerning propagation of the vortex beam have been considered in the literature so far. Some of them consider optical fields containing the lattice of vortices generated for example by three or more plane or spherical waves interference \cite{6:Masajada,7:Senthilkumaran,8:Ruben,9:Vyas}; or so called composed vortices, i.e. vortices which are generated by two or more overlapping beams \cite{10:Vasnetsov,11:Bekshaev,12:Yang,13:Khonina,14:Bouchal,15:Sokolenko,16:Szatkowski}.\par
The study on single vortex beam propagation has started with the most basic problem, that is propagation of the fundamental vortex beam in a free space \cite{17:Bazhenov,18:Indebetouw,19:Rozas}. Next, the problem of vortex beam propagation through the simple optical element revealing circular symmetry was reported \cite{20:Topuzoski,21:Khonina}. Here, the most important part for our considerations are papers devoted to Gaussian beam propagation through the spiral phase plate (SPP) \cite{22:Beijersbergen,23:Allen,24:Swartzlander,25:Khoroshun,26:Kotlyar,27:Khonina}. SPP is now one of the most common ways of introducing optical vortex into the laser beam. In more advanced approaches the propagation of vortex beam with broken symmetry or through the system with broken symmetry (like for example diffraction by half-plane \cite{28:Masajada,29:Bekshaev} or a phase step \cite{30:Masajada,31:Spektor}) was studied. Most of these asymmetrical cases were studied combining numerical and/or strongly approximated analytical method, especially in case of higher order vortices \cite{32:Masajada,33:Anzolin,34:Singh,35:Singh,36:Masajada,37:Dennis,38:Augustyniak,39:Popiolek,40:Savelyev}. Another highly asymmetrical problem is a vortex beam propagation through the turbid media (e.g. atmosphere) \cite{41:Aksenov,42:Morgan,43:Soifer}.\par

\begin{figure}
\includegraphics[scale=0.48]{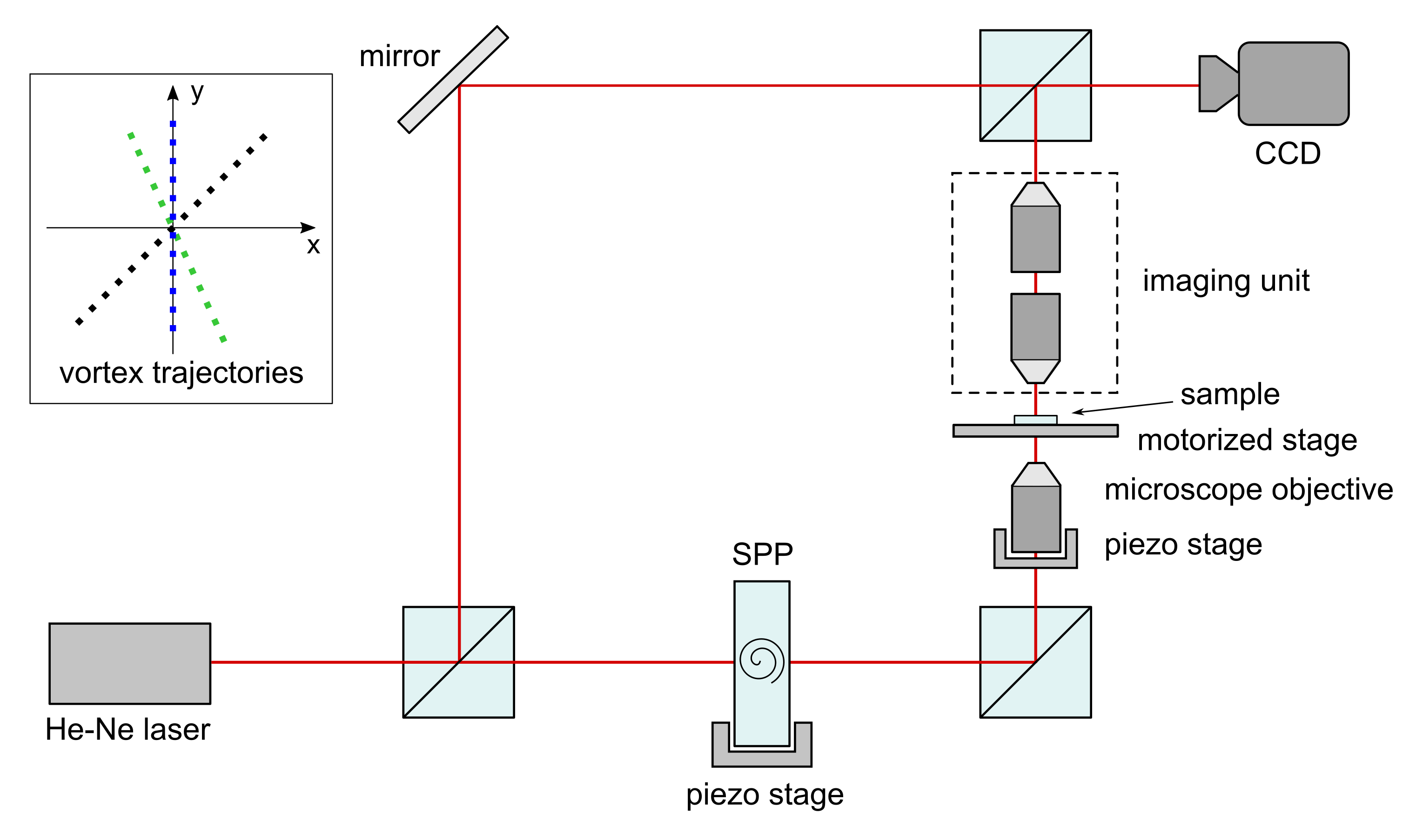}
\caption{The scheme of optical vortex scanning microscope (OVSM). In the inset the registered three vortex trajectories are shown. The vortex trajectories are subsequent positions of vortex point (vortex point is a point where the phase is singular) at sample plane while moving the SPP perpendicularly to the optical beam. The three trajectories were measured at three slightly different positions of focusing objective. This objective was moved toward the sample.  \label{fig:experimental_setup}}
\end{figure}

In the paper \cite{33:Anzolin} the analysis of the Gaussian beam propagation through the off-axis SPP in Fraunhofer approximation is studied. Authors have focused their attention on the vortex point displacement measured by inspecting the asymmetry in intensity distribution at the far field. In the present paper we analyze asymmetrical optical system with Fresnel diffraction theory, which is more general than Fraunhofer one. The analysis of the off-axis high-order vortices is a difficult task. The integrals become highly complicated and some typical tricks often used in the calculations cannot be applied. The good example is stationary phase method \cite{44:Stamnes}, which cannot be used since the phase changes very fast in the vicinity of the vortex point. That is the reason why there are only few publications regarding the exact solutions of asymmetric higher order vortex propagation. In paper \cite{45:Kotlyar} the elliptic vortex beam propagation is studied. The paper \cite{46:Rumi} describes the generation of the higher order vortex beam by discretizing spiral phase plate. In paper \cite{47:Berry} the generation of vortex beam through fractional spiral phase plate is studied. In papers \cite{48:Bekshaev,49:Bekshaev} the propagation through off-axis hologram generating the optical vortices is analyzed, also including the effects of misalignment. In papers \cite{50:Plocinniczak,51:Plociniczak} we have provided a solution for asymmetrical vortex propagation in optical vortex scanning microscope (OVSM)  presented schematically in Fig.~\ref{fig:experimental_setup}. In this paper we propose more general solution in terms of Kummer confluent hypergeometric function which can be used for a system of arbitrary number of lenses. \par

\begin{figure}
\includegraphics[scale=0.12]{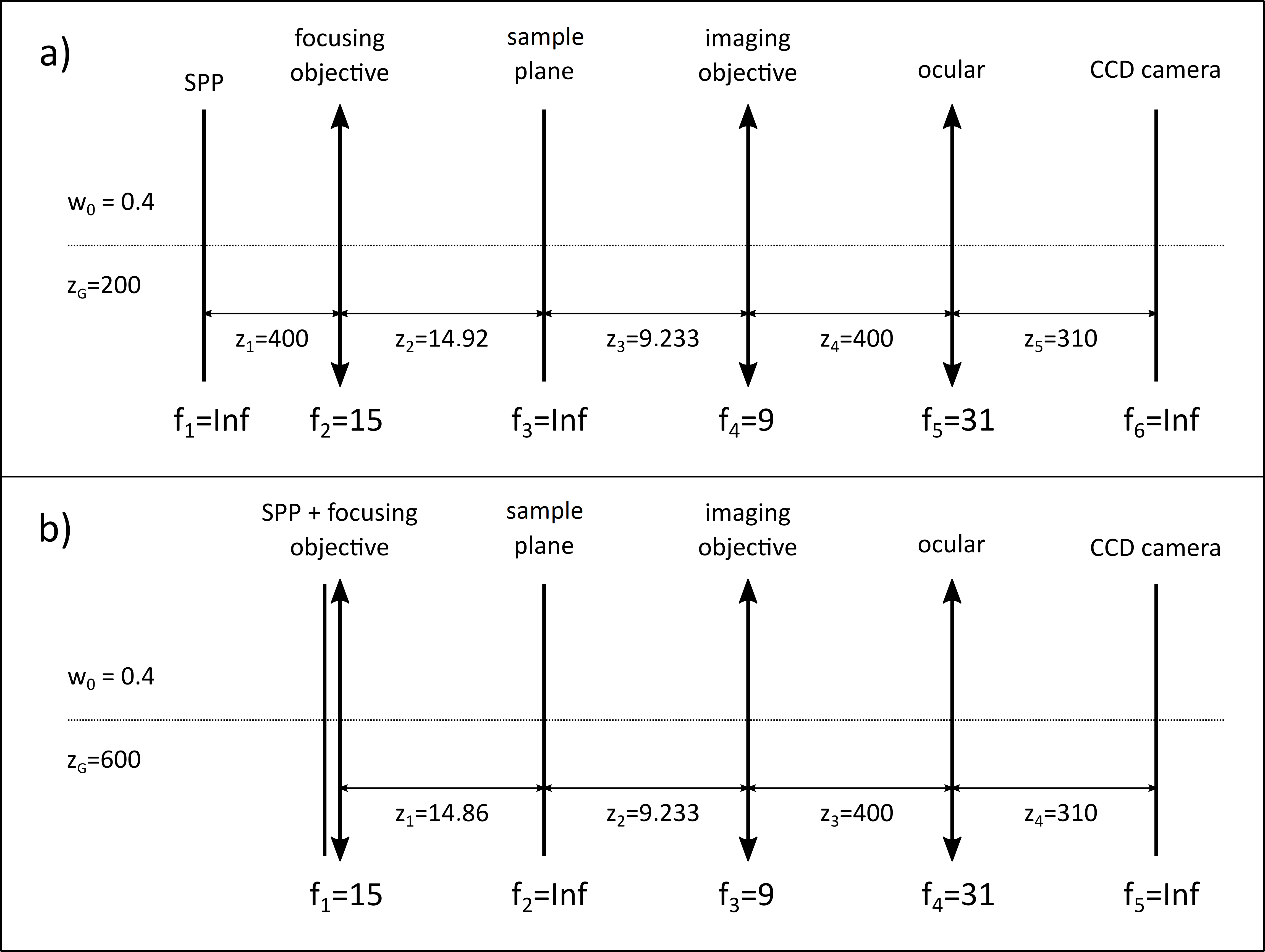}
\caption{Two versions of the OVSM scheme; a) the SPP is separated from the focusing lens; b) the SPP and focusing objective work together as a single thin element. In both cases the imaging objective and the ocular form the image of the sample plane on CCD camera. Units: [mm].\label{fig:two_setups_ab}}
\end{figure}

In this paper the optical system shown schematically in Fig.~\ref{fig:two_setups_ab} is studied. The system represents the object arm of the OVSM shown in Fig.~\ref{fig:experimental_setup} (in experiment, the reference arm is necessary for reconstructing both amplitude and phase of the object beam. Obviously, in analytical and numerical calculations we do not need it). The system is divided into blocks. Each block consists of one or more elements and is represented by its transmittance function (in case of single element) or the product of the transmittance functions (when there are two or more elements in the given block). In our first approach \cite{50:Plocinniczak} the OVSM was reduced to a single block consisting of three elements, i.e. incident Gaussian beam, SPP and focusing lens considered as a single thin element. The image was calculated at the sample plane (noted as sample in Fig.~\ref{fig:experimental_setup}). It should be noticed that the SPP can be moved perpendicularly to the optical axis, which breaks the system symmetry. In result the vortex point moves inside the focused beam, but the range of this movement is highly reduced due to focusing lens. The inset in Fig.~\ref{fig:experimental_setup} shows the exemplary vortex trajectories as registered in our experimental system. In this way the sample can be scanned with the vortex point (i.e. point where the phase is singular). This technique is named the Internal Scanning Method (ISM) \cite{36:Masajada,38:Augustyniak,50:Plocinniczak,51:Plociniczak,52:Popiolek,53:Popiolek}. In the paper \cite{51:Plociniczak} the system built of three blocks was analyzed. The first block consisted of incident Gaussian beam, SPP and focusing lens, the second was just the sample plane, and the third contained a single imaging lens. Here, we extend the analysis to the fully expanded system shown in Fig.~\ref{fig:two_setups_ab}(b).

Our analysis was performed within the frame of scalar diffraction theory in Fresnel approximation \cite{54:Goodman}. The Fresnel diffraction integral for the first block shown in Fig.~\ref{fig:two_setups_ab}(b) has a form

\begin{equation}
\begin{split}
u(x_1,y_1) & = \iint \displaylimits_{\mathbb{R}^2} T_{\nu}(x,y)\, u_G(x-x_c,y)\, T_f(x-x_c,y)\, \\ & \times e^{\frac{ik}{2z_1}(x_1^2+y_1^2)}\, e^{\frac{ik}{2z_1}[(x-x_c)^2+y^2]}\, e^{-\frac{ik}{z_1}[(x-x_c)x_1+yy_1]} \mathrm{d}x \mathrm{d}y
\end{split}
\label{eq:1}
\end{equation}

where $T_v$ is transmittance function of the SPP, $u_G$ is an incident Gaussian beam, $T_f$ is a transmittance function of the focusing lens.

\begin{subequations}
\begin{equation}
T_{\nu} = e^{im\phi}
\label{eq:2a}
\end{equation}
\begin{equation}
u_G = e^{-\left(\tfrac{1}{w^2(z)}+\tfrac{ik}{2R(z)}\right) [(x-x_c)^2+y^2]}
\label{eq:2b}
\end{equation}
\begin{equation}
T_f = e^{-\frac{ik}{2f_1} [(x-x_c)^2+y^2]}
\label{eq:2c}
\end{equation}
\begin{equation}
w^2(z) = w_0^2\left(1+\left(\frac{\lambda z}{\pi w_0^2}\right)^2\right)
\label{eq:2d}
\end{equation}
\begin{equation}
R(z) = z\left(1+\left(\frac{\pi w_0^2}{\lambda z}\right)^2\right)
\label{eq:2e}
\end{equation}
\end{subequations}

$m$ is a topological charge of the optical vortex; it is an integer -- positive or negative, $w_0$ is a beam waist (of the Gaussian beam), $f_1$ is a focal length of the focusing lens.\par

Instead of moving the SPP off the optical axis by the distance $x_c$ we moved the rest of the system (including the screen) by the same $x_c$, which simplified further calculations. As was shown in \cite{50:Plocinniczak} the integral Eq.~(\ref{eq:1}) had a solution

\begin{equation}
u_1(x_1,y_1) = \Xi_1 K(A^{(1)},B_x^{(1)},B_y^{(1)},C^{(1)})
\label{eq:3}
\end{equation}

where
\begin{subequations}
\begin{equation}
\Xi_1 = \frac{e^{ikz_1}}{i \lambda z_1}\, e^{iknd_1} U_0\, \frac{w_0}{w(z)} e^{i[\arctan(\frac{z}{z_R})-kz+\omega t]}
\label{eq:4a}
\end{equation}
\begin{equation}
A^{(1)} = \alpha+i\beta
\label{eq:4b}
\end{equation}
\begin{equation}
B_x^{(1)} = -2x_c (\alpha+i\beta)-\frac{ikx_1}{z_1}
\label{eq:4c}
\end{equation}
\begin{equation}
B_y^{(1)} = - \frac{iky_1}{z_1}
\label{eq:4d}
\end{equation}
\begin{equation}
C^{(1)} = x_c^2 (\alpha+i\beta)+\frac{ik}{2z_1} (x_1^2+y_1^2)+\frac{ikx_c x_1}{z_1}
\label{eq:4e}
\end{equation}

The parameters $\alpha$ and $\beta$ are
\begin{equation}
\alpha = -\frac{1}{w^2(z)}
\label{eq:4f}
\end{equation}
\begin{equation}
\beta = -\frac{k}{2} \left(\frac{1}{R(z)}+\frac{1}{f_1}-\frac{1}{z_1}\right)
\label{eq:4g}
\end{equation}
\label{eq:4}
\end{subequations}

The main part of this solution is function Kappa $K$.

\begin{subequations}
\begin{equation}
\begin{split}
K(A,B_x,B_y,C) = -\frac{\sqrt{\pi} \cdot e^C}{2A} \\ \times
\begin{cases} \sqrt{\pi}\sum_{n=0}^\infty \frac{1}{n!(2\sqrt{-A})^{2n+1}} \sum_{j=0}^{2n+1} {{2n+1}\choose{j}} B_x^j B_y^{2n+1-j} \sum_{l=0}^m {{m}\choose{l}} i^l J_{m+j-l,2n+1-j+l}; \, \mathrm{for}\, m\,\mathrm{odd} \\ \sum_{n=0}^\infty \frac{2^{n+1}}{(2n+1)!!(2\sqrt{-A})^{2n+2}} \sum_{j=0}^{2n+2} {{2n+2}\choose{j}} B_x^j B_y^{2n+2-j} \sum_{l=0}^m {{m}\choose{l}} i^l J_{m+j-l,2n+2-j+l}; \, \mathrm{for}\, m\,\mathrm{even} \end{cases}
\end{split}
\label{eq:5a}
\end{equation}
where
\begin{equation}
J_{\delta, \eta} = \int_{0}^{2\pi} (\cos\varphi)^\delta (\sin\varphi)^\eta \mathrm{d}\varphi
\label{eq:5b}
\end{equation}
\label{eq:5}
\end{subequations}

We have obtained an interesting result showing that a system having more blocks (with lenses or just planes) can still be described by Kappa function, but with different coefficients $\Xi_q$, $A^{(q)}$, $B_x^{(q)}$, $B_y^{(q)}$, $C^{(q)}$. The number of blocks in the system is indicated by superscript ($q$) as it has been already done in Eq.~(\ref{eq:3}) and Eq.~(\ref{eq:4}). In paper \cite{51:Plociniczak} we calculated explicitly the coefficients $\Xi_q$, $A^{(q)}$, $B_x^{(q)}$, $B_y^{(q)}$, $C^{(q)}$ up to $q=3$, and postulated that it can be done for any ($q$). In this paper we derived a recurrence formula for the coefficient for any number of blocks (any ($q$)), which is actually a formal proof of our previous claim.\par 
The function Kappa was denoted previously as $G$ \cite{50:Plocinniczak,51:Plociniczak}. Since that time we have improved and generalized our solution. Now the function $G$ is rewritten in more useful form. In order not to confuse both versions, we denote the present version by capital Kappa $K$. \par
\begin{figure}[h!]
\includegraphics[scale=0.4]{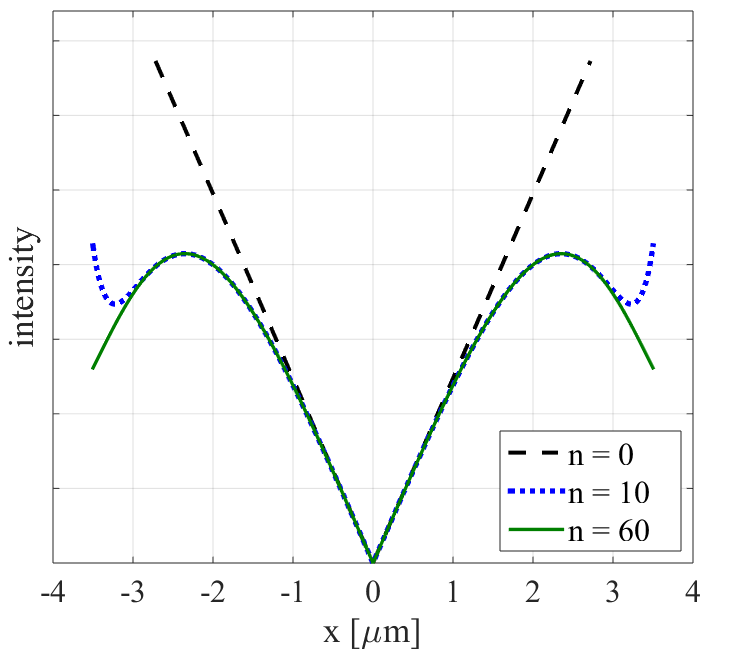}
\caption{The amplitude distribution of the vortex beam for the system shown in Fig.~\ref{fig:two_setups_ab}(b) in case of $n=0$, $n=10$ and $n=60$. \label{fig:vortex_crossection}}
\end{figure}

As it was shown in papers \cite{50:Plocinniczak,51:Plociniczak}, close to the vortex axis the $n=0$ term is sufficient to evaluate the vortex beam. Figure~\ref{fig:vortex_crossection} illustrates this fact, but now it is plotted for the four-block system shown in Fig.~\ref{fig:two_setups_ab}(b). This is a useful result for us, as we analyze the OVSM images at the central part of the beam, where the $n=0$ approximation well represents phase and amplitude distribution of our beam. 

Our goal was to represent the entire object arm of the OVSM as presented in Fig.~\ref{fig:two_setups_ab}(a). Now the SPP can be separated from the focusing lens which means that focal length for the first block (i.e. SPP block) is \mbox{$f_1=\infty$}. Next there is a focusing lens with focal length \mbox{$f_2=15$~mm} followed by the sample plane for which the focal length is also infinite \mbox{$f_3=\infty$}. The third block is the imaging lens with focal length \mbox{$f_4=9$~mm} and the last one is the ocular lens with focal length \mbox{$f_5=31$~mm}, after which we have observation plane (screen). Unfortunately, for the reasons given later in the manuscript the full system could not be analyzed with Kappa function. The SPP cannot be separated from the focusing lens. Therefore, we have to switch to the system shown in Fig.~\ref{fig:two_setups_ab}(b). In this paper we will analyze this system showing the efficiency of our formulas. As can be noticed from Fig.~\ref{fig:two_setups_ab}(b), in our model a sample plane is treated as a separate block. In this way we are able to enhance our calculations in order to analyze the influence of a simple phase object on the vortex beam. Thus, the system shown in Fig.~\ref{fig:two_setups_ab}(b) prepares us for this next step.\par

Certainly, all the blocks may have different focal lengths and positions than the ones shown in Fig.~\ref{fig:two_setups_ab} and the Kappa function will still work. However, the important thing is that the first element must be the block containing the SPP, focusing lens and Gaussian beam.\par
The present paper is organized as follows. In section \ref{sect:analitical_part} the extension of our formulas as well as their closed form for the system consisting of any number of lenses is discussed. In section~\ref{sect:coeff_A} we discussed the role of coefficient $A$. In section~\ref{sect:ice-skater_effect} we show the efficiency of our formulas by discussing the effect of vortex trajectory rotation. Section~\ref{sect:conclusions} concludes the paper.

\section{Analytical part \label{sect:analitical_part}}
In this part our previous results will be extended. In order to build more relevant model of the OVSM we have derived the explicit formula for coefficients $\Xi_q$, $A^{(q)}$, $B_x^{(q)}$, $B_y^{(q)}$, $C^{(q)}$ up to $q=4$, just to analyze the OVSM system shown in Fig.~\ref{fig:two_setups_ab}(b).

\begin{subequations}
\begin{equation}
\Xi_4 = \frac{e^{ik(z_1+z_2+z_3+z_4)}}{\lambda^4 z_1z_2z_3z_4}\, e^{ikn(d_1+d_2+d_3+d_4)} U_0\, \frac{w_0}{w(z)} e^{i[\arctan(\frac{z}{z_R})-kz+\omega t]}
\label{eq:6a}
\end{equation}
\begin{equation}
A^{(4)} = \alpha+i\beta-\frac{ik}{2z_1^2\gamma_2}-\frac{ik}{2z_1^2z_2^2\gamma_2^2\gamma_3}-\frac{ik}{2z_1^2z_2^2z_3^2\gamma_2^2\gamma_3^2\gamma_4}
\label{eq:6b}
\end{equation}
\begin{equation}
B_x^{(4)} = -2x_c (\alpha+i\beta)+\frac{ikx_c}{z_1^2\gamma_2}+\frac{ikx_c}{z_1^2z_2^2\gamma_2^2\gamma_3}-\frac{ik}{z_1z_2z_3\gamma_2\gamma_3\gamma_4}\, \left(\frac{x_4}{z_4}-\frac{x_c}{z_1z_2z_3\gamma_2\gamma_3}\right)
\label{eq:6c}
\end{equation}
\begin{equation}
B_y^{(4)} = -\frac{iky_4}{z_1z_2z_3z_4\gamma_2\gamma_3\gamma_4}
\label{eq:6d}
\end{equation}
\begin{equation}
\begin{split}
C^{(4)} & = x_c^2 (\alpha+i\beta)+\frac{ik}{2z_4} (x_4^2+y_4^2)-\frac{ikx_c^2}{2z_1^2\gamma_2}-\frac{ikx_c^2}{2z_1^2z_2^2\gamma_2^2\gamma_3} \\ & -\frac{ik}{2\gamma_4}\, \left[\left(\frac{x_4}{z_4}-\frac{x_c}{z_1z_2z_3\gamma_2\gamma_3}\right)^2+\left(\frac{y_4}{z_4}\right)^2\right]
\end{split}
\label{eq:6e}
\end{equation}
\begin{equation}
\gamma_4 = \frac{1}{z_3}+\frac{1}{z_4}-\frac{1}{f_4}-\frac{1}{z_3^2\gamma_3}
\label{eq:6f}
\end{equation}
\begin{equation}
\gamma_3 = \frac{1}{z_2}+\frac{1}{z_3}-\frac{1}{f_3}-\frac{1}{z_2^2\gamma_2}
\label{eq:6g}
\end{equation}
\begin{equation}
\gamma_2 = \frac{1}{z_1}+\frac{1}{z_2}-\frac{1}{f_2}
\label{eq:6h}
\end{equation}
\end{subequations}

where the meaning of $z_4$ and $f_4$ can be read from Fig.~\ref{fig:two_setups_ab}), $d_1$,…, $d_4$ are lens thicknesses. In case of a simple plane we put $d=0$.\par
Following this path a general iterative formulas for coefficient of any ($q$) index can be derived. The formulas have a form

\begin{subequations}
\begin{equation}
\begin{split}
u_j(x_j,y_j) & = \xi_j \iint \displaylimits_{\mathbb{R}^2} u_{j-1}(x_{j-1},y_{j-1})\, e^{\frac{ik}{2z_j}(x_j^2+y_j^2)}\, e^{\frac{ik}{2z_j}(x_{j-1}^2+y_{j-1}^2)}\, e^{-\frac{ik}{2f_j}(x_{j-1}^2+y_{j-1}^2)}\, \\ & \times e^{-\frac{ik}{z_j}(x_{j-1}x_j+y_{j-1}y_j)} \mathrm{d}x_{j-1} \mathrm{d}y_{j-1}
\end{split}
\label{eq:7a}
\end{equation}
\begin{equation}
u_j(x_j,y_j) = \Xi_j \left(\frac{2i\pi}{k}\right)^{j-1} \prod_{s=1}^j \frac{1}{\gamma_s} \cdot K(A^{(j)},B_x^{(j)},B_y^{(j)},C^{(j)})
\label{eq:7b}
\end{equation}
\end{subequations}

\begin{subequations}
\begin{equation}
A^{(j)} = \alpha+i\beta-\frac{ik}{2}\sum_{p=1}^{j-1}\prod_{s=1}^p \frac{1}{z_s^2\gamma_s^2\gamma_j}
\label{eq:8a}
\end{equation}
\begin{equation}
B_x^{(j)} = -2x_c (\alpha+i\beta)+ikx_c\sum_{p=1}^{j-1}\prod_{s=1}^p \frac{1}{z_s^2\gamma_s^2\gamma_j}-\frac{ikx_j}{\prod_{s=1}^j z_s\gamma_s}
\label{eq:8b}
\end{equation}
\begin{equation}
B_y^{(j)} = -\frac{iky_j}{\prod_{s=1}^j z_s\gamma_s}
\label{eq:8c}
\end{equation}
\begin{equation}
C^{(j)} = x_c^2 (\alpha+i\beta)+\frac{ik}{2z_j} (x_j^2+y_j^2)-\frac{ikx_c^2}{2} \sum_{p=1}^{j-1}\prod_{s=1}^p \frac{1}{z_s^2\gamma_s^2\gamma_j}+\frac{ikx_cx_j}{2\prod_{s=1}^j z_s\gamma_s}-\frac{ik}{2z_j^2\gamma_j}(x_j^2+y_j^2)
\label{eq:8d}
\end{equation}
\begin{equation}
\gamma_j = \frac{1}{z_{j-1}}+\frac{1}{z_j}-\frac{1}{f_j}-\frac{1}{z_{j-1}^2\gamma_{j-1}}
\label{eq:8e}
\end{equation}
\begin{equation}
\Xi_j = \frac{e^{ik\sum_{s=1}^j z_s}}{i^j\lambda^j \prod_{s=1}^j z_s}\, e^{ikn\sum_{s=1}^j d_s} U_0\, \frac{w_0}{w(z)} e^{i[\arctan(\frac{z}{z_R})-kz+\omega t]}
\label{eq:8f}
\end{equation}
\begin{equation}
\gamma_0 = \gamma_1 = 1,\, j\in \mathbb{N}_+
\label{eq:8g}
\end{equation}
\label{eq:8}
\end{subequations}

The derivation of the above is presented in Appendix~\ref{appendix_A}.\\
The sums in Kappa function are convergent provided that a series of conditions are hold

\begin{equation}
\frac{1}{z_j}>\frac{1}{f_j}+\frac{1}{\gamma_{j-1}}-\frac{1}{z_{j-1}}; \, j\leq q
\label{eq:9}
\end{equation}

Very similar formulas can be derived for negative vortex charge. In that case we can use the same Kappa function but with some multiplying expression and $B_y$ coefficient multiplied by $-1$.

\begin{equation}
u_{j-}(x_j,y_j) = -e^{-i2\pi m}\, \Xi_j\, K(A^{(j)},B_x^{(j)},-B_y^{(j)},C^{(j)})
\label{eq:10}
\end{equation}

The derivation of this formula is presented in Appendix~\ref{appendix_B}. \\
We have also derived a closed formula, but with a special function, which is given below. The derivation is discussed in the Appendix~\ref{appendix_C}.

\begin{subequations}
\begin{equation}
\begin{split}
u_q(x_q,y_q) & = -\Xi_q \frac{\pi\sqrt{\pi}e^{C^{(q)}}}{2^m\sqrt{-A^{(q)}}} \frac{1}{\left(\frac{m-1}{2}\right)!}\, \left(\frac{B_x^{(q)}}{2\sqrt{-A^{(q)}}}+i\frac{B_y^{(q)}}{2\sqrt{-A^{(q)}}}\right)^m \\ & \times _1F_1\left(1+\frac{m}{2},1+m,\left(\frac{B_x^{(q)}}{2\sqrt{-A^{(q)}}}\right)^2+\left(\frac{B_y^{(q)}}{2\sqrt{-A^{(q)}}}\right)^2\right); \mathrm{for}\, m\,\mathrm{odd}
\end{split}
\label{eq:11a}
\end{equation}
\begin{equation}
\begin{split}
u_q(x_q,y_q) & = \Xi_q \frac{\pi e^{C^{(q)}}}{2^{\frac{m}{2}}\sqrt{-A^{(q)}}} \frac{1}{(m-1)!!}\, \left(\frac{B_x^{(q)}}{2\sqrt{-A^{(q)}}}+i\frac{B_y^{(q)}}{2\sqrt{-A^{(q)}}}\right)^m \\ & \times _1F_1\left(1+\frac{m}{2},1+m,\left(\frac{B_x^{(q)}}{2\sqrt{-A^{(q)}}}\right)^2+\left(\frac{B_y^{(q)}}{2\sqrt{-A^{(q)}}}\right)^2\right); \mathrm{for}\, m\,\mathrm{even}
\end{split}
\label{eq:11b}
\end{equation}
\label{eq:11}
\end{subequations}

where $_1F_1$ is the Kummer confluent hypergeometric function \cite{55:Abramowitz}. The calculations using closed formula are much faster, but as we will show both formulas Eq.~(\ref{eq:3}) and Eq.~(\ref{eq:11}) are helpful in understanding the vortex beam propagation through our system. \par
The sum of two $B$ coefficients in Eq.~(\ref{eq:8b})-(\ref{eq:8c}) is a complex expression which can be easily decomposed into real and imaginary part. From Eq.~(\ref{eq:8b})-(\ref{eq:8c}) we got for real part

\begin{equation}
-2x_c\alpha+\frac{ky_q}{\xi_a^{(q)}} \Rightarrow y_q=2x_c\frac{\alpha}{k}\xi_a^{(q)}
\label{eq:12}
\end{equation}

And for imaginary part

\begin{equation}
x_c\, (-2\beta+k\xi_b^{(q)})+k\xi_c^{(q)}x_q \Rightarrow x_q=-x_c\, \frac{(-2\beta+k\xi_b^{(q)})}{k\xi_c^{(q)}}
\label{eq:13}
\end{equation}

The form of coefficients $\xi_a^{(q)}$, $\xi_b^{(q)}$, $\xi_c^{(q)}$ depends on the value of $q$.
 
\begin{subequations}
\begin{equation}
\xi_a^{(q)} = z_1z_2z_3 \dots z_q\gamma_2 \dots \gamma_q
\label{eq:14a}
\end{equation}
\begin{equation}
\xi_b^{(1)} = 0,\, \mathrm{otherwise}\; \xi_b^{(q)} = \frac{1}{z_1^2\gamma_2}+\frac{1}{z_1^2z_2^2\gamma_2^2\gamma_3}+ \dots + \frac{1}{z_1^2 \dots z_{q-1}^2\gamma_2^2 \dots \gamma_{q-1}^2\gamma_q}
\label{eq:14b}
\end{equation}
\begin{equation}
\xi_c^{(q)} = \frac{1}{z_1 \dots z_q\gamma_2 \dots \gamma_q}
\label{eq:14c}
\end{equation}
\label{eq:14}
\end{subequations}

For example for $q=4$ we get

\begin{subequations}
\begin{equation}
\xi_a^{(4)} = z_1z_2z_3z_4\gamma_2\gamma_3\gamma_4
\label{eq:15a}
\end{equation}
\begin{equation}
\xi_b^{(4)} = \frac{1}{z_1^2\gamma_2}+\frac{1}{z_1^2z_2^2\gamma_2^2\gamma_3}+\frac{1}{z_1^2z_2^2z_3^2\gamma_2^2\gamma_3^2\gamma_4}
\label{eq:15b}
\end{equation}
\begin{equation}
\xi_c^{(4)} = \frac{1}{z_1z_2z_3z_4\gamma_2\gamma_3\gamma_4}
\label{eq:15c}
\end{equation}
\end{subequations}

Formulas in Eq.~(\ref{eq:12}) and Eq.~(\ref{eq:13}) show that y-coordinate is a member of imaginary part and x-coordinate is a member of real part of the $B_x+iB_y$ expression. This leads us to simple formulas for vortex point trajectory. The vortex point (zero amplitude point) is at the place where both imaginary and real part are equal to zero \cite{1:Vasnetsov,2:Soskin,3:Dennis}.
 
\begin{subequations}
\begin{equation}
-2x_c\alpha+\frac{ky_q}{\xi_a^{(q)}} = 0 \Rightarrow y_q=2x_c\frac{\alpha}{k}\xi_a^{(q)}
\label{eq:16a}
\end{equation}
\begin{equation}
x_c\, (-2\beta+k\xi_b^{(q)})+k\xi_c^{(q)}x_q = 0 \Rightarrow x_q=-x_c\, \frac{(-2\beta+k\xi_b^{(q)})}{k\xi_c^{(q)}}
\label{eq:16b}
\end{equation}
\label{eq:16}
\end{subequations}

From the above formula we may conclude that the vortex trajectory as a function of $x_c$, for the given z is a straight line. We may also find a plane where the vortex trajectory is perpendicular to the SPP shift. Using Eq.~(\ref{eq:16b}) the condition is

\begin{equation}
x_q = 0 \Rightarrow -2\beta+k\xi_b^{(q)} = 0
\label{eq:17}
\end{equation}

In the paper \cite{50:Plocinniczak} we have formulated the hypothesis that the higher order vortices ($m>1$) do not split even when $x_c\neq 0$. The higher order vortices are classified as structurally unstable \cite{56:Nye}, i.e. they are supposed to split into single order vortices even under small phase or amplitude perturbation \cite{57:Ricci,58:Neo,59:Zambrana}. Nevertheless, the formulas in Eq.~(\ref{eq:11}) proof the stability hypothesis in an explicit form. The factors in front of the Kummer confluent hypergeometric function are just a vortex term.

\begin{equation}
\left(\frac{B_x^{(q)}}{2\sqrt{-A^{(q)}}}+i\frac{B_y^{(q)}}{2\sqrt{-A^{(q)}}}\right)^m
\label{eq:18}
\end{equation}

\begin{figure}
\includegraphics[scale=0.48]{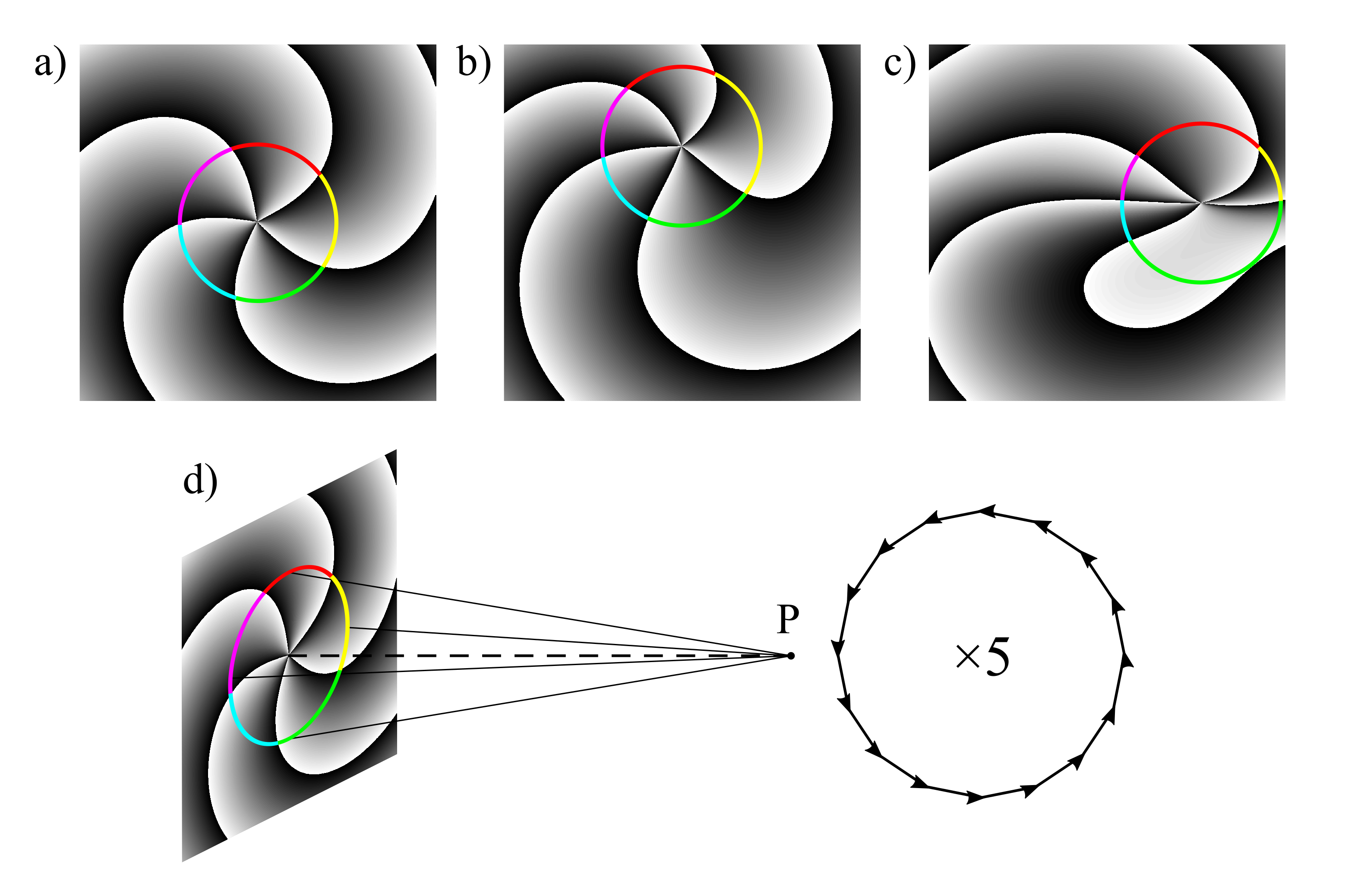}
\caption{The phase distribution of the optical vortex beam ($m=5$) for $x_c=0$ as seen just behind the SPP. Part d shows the sum of phasors calculated at point P (at image plane) on the optical axis, while going along the colored circle. Due to symmetry and 2$\pi$ phase change along the green part of the circle the phasors (from the green part) form a full circle at sum up to zero. Going along the parts marked with different colors will produce five such circles of phasors, at the same place, which again sum to zero; b) when $x_c\neq 0$, the symmetry breaks. In a typical case each part of the circle noted by different color will form zero at different point at the image plane, splitting the $m$-th order vortex into a set of single vortices. But our case is untypical and the higher order vortex does not split; c) when performing more complex operation as described below, we will produce even less symmetrical phase distribution, but the higher order vortex is still stable. \label{fig:phase_maps_circles}}
\end{figure}

The place where real and imaginary part of $B_x^{(q)}$+$iB_y^{(q)}$ is zero indicates the position of the vortex point of the $m$-th order. Since the whole term is at power $m$, the $m$-order vortex does not split for any $x_c$. The result is very interesting, which is illustrated in Fig.~\ref{fig:phase_maps_circles}.

\begin{figure}
\includegraphics[scale=0.9]{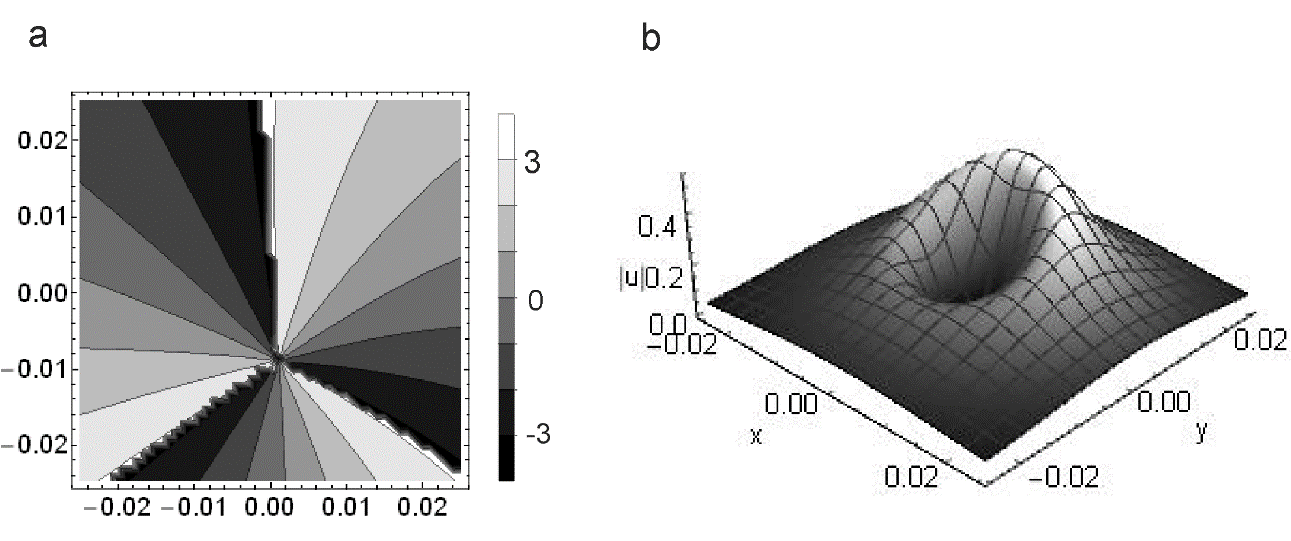}
\caption{The vortex beam of $m=3$ focused by the lens $f_1=15$ mm (according to Fig.~\ref{fig:two_setups_ab}(a)); a) phase and b) amplitude distribution, calculated at $z=14.88$ mm. Both distributions are evidently asymmetrical ($x_c=0.15$ mm). Nevertheless, they do not split while propagating through the rest of the OVSM system. \label{fig:asymmetrical_vortex}}
\end{figure}

The situation would not be strange if the higher order optical vortex were  stable for a given non-zero value $x_c$. However, changing the $x_c$ continuously breaks the circular symmetry of the phase distribution at the SPP plane with no harm for the higher order OV stability. We could also start our analysis at the sample plane, where both phase and amplitude distribution symmetry is broken (Fig.~\ref{fig:asymmetrical_vortex}). Nevertheless, the propagation of this input beam through any number of lenses will not split the higher order vortices.\par

\begin{figure}
\includegraphics[scale=0.9]{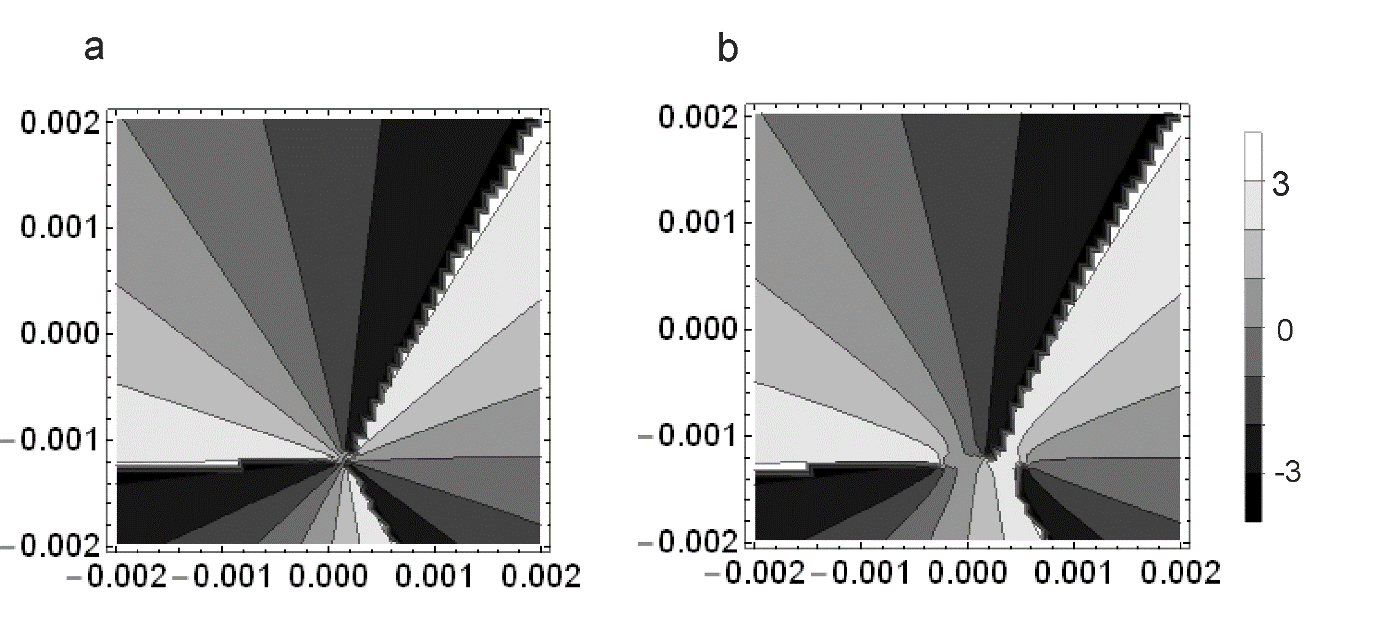}
\caption{The phase image of focused vortex beam (parameters as in Fig.~\ref{fig:two_setups_ab}(b)), $m=3$; a) calculated from Kappa function with $n=21$. The higher order vortex is still there; b) calculated from Kappa function with $n=21$, but the number $0.05$ was added to binomial factor in the second sum for $n=1$. The vortex was split into three single vortices. \label{fig:split_vortex}}
\end{figure}

We can go even further. Adding the term inside the bracket in Eq.~(\ref{eq:18}) and multiplying $B_x$ or $B_y$ coefficients by any number, but in such a way that the coefficients by $x_{(q)}$ remain real (or become imaginary) and coefficient by $y_{(q)}$ remain imaginary (or become real) may change the position of the vortex point and its phase distribution (Fig.~\ref{fig:asymmetrical_vortex}(c)) but still the higher order vortices will be stable while propagating through our system. However, changing other parameters, like for example binomial factor by coefficients $B$ in Eq.~(\ref{eq:5}) splits the higher order vortices immediately (Fig.~\ref{fig:split_vortex}).  

To summarize this part: We have shown that the higher order vortices when propagating through the set of classical lenses described in paraxial approximation will not split regardless of asymmetry introduced by the off axis position of the SPP. Moreover, we may perform some more symmetry breaking operation (as shown in Fig.~\ref{fig:phase_maps_circles}(c)) in Eq.~(\ref{eq:8c})-(\ref{eq:8d}), provided that the coefficients by $x$ remain real (or imaginary) and by $y$ imaginary (or real). So, the conclusion is that the very basic optical system does not split the higher order vortices even if the input phase and amplitude distribution is highly non-symmetrical. In many cases such an unusual stability results from deeper physical rules. There is a question if this is also a case here. So far we have no answer to this question. What we can learn now is that classical optical system is somewhat special, at least when being described in paraxial approximation and illuminated by Gaussian beam with the vortex beam introduced by SPP. In the next section the special role of the coefficient $A$ will be studied. 

\section{Coefficient A\label{sect:coeff_A}}

In our further discussion we will refer to two specific examples of the OVSM models. One with the separated SPP plate and focusing lens (Fig.~\ref{fig:two_setups_ab}(a)) and the second one with SPP plate and focusing lens working as single thin element (Fig.~\ref{fig:two_setups_ab}(b)).\par 
The coefficient $A^{(q)}$ has relatively simple form. It depends neither on the $x$ and $y$ coordinates nor on SPP shift $x_c$. However, it plays a crucial role in vortex beam phase evolution. Unfortunately it cannot be totally taken outside the first sum in Kappa function. On the contrary, in Eq.~(\ref{eq:4}) it can be entirely assimilated inside the first sum, but in the present form some mathematical aspects can be noticed in a more clear way. \par
\begin{figure}
\includegraphics[scale=0.9]{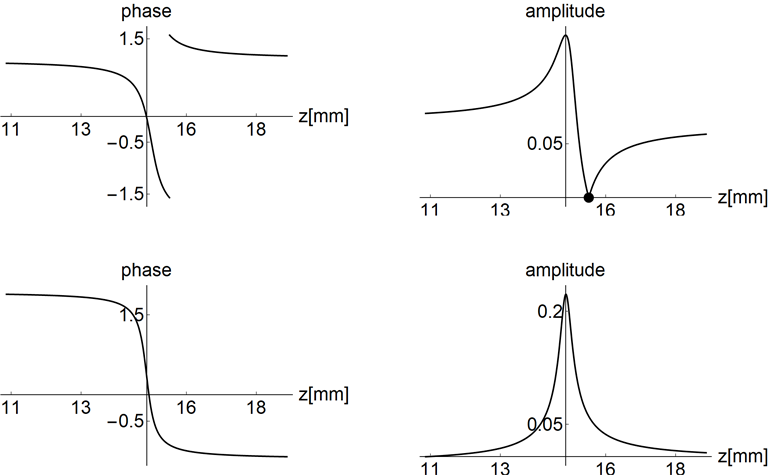}
\caption{The phase (a) and amplitude (b) distribution of the $1/A^{(2)}$ in case of system with the separated SPP and focusing lens as a function of focusing lens position in respect to the sample plane. In (c) and (d) the phase and amplitude distribution of the $1/A^{(1)}$ term is shown in case of SPP and focusing lens working as a single element (with no separation). \label{fig:coeff_A_plots}}
\end{figure}

The first important point is that the coefficient $A$ is responsible for breaking the Kappa function convergence. In the OVSM this happens when the SPP is separated from the focusing lens, and coefficient $A$ has a singular point (Fig.~\ref{fig:coeff_A_plots}(a,b)). At first we will analyze the system with two blocks, the SPP and the focusing lens (sample plane is an observation plane in this case), so we need the second order coefficient $A^{(2)}$. The coefficient $A^{(q)}$ contributes to the Kappa function as $1/A^{(q)}$, so the term in front of the first sum takes form (for $q=2$). \par

\begin{equation}
\frac{1}{A^{(2)}} = \left(\alpha+i\beta-\frac{ik}{2z_1^2\gamma_2}\right)^{-1}
\label{eq:19}
\end{equation}

There exist a range of such positive $z_2$ for which the conditions in Eq.~(\ref{eq:9}) fail. In particular there is a $z_2$ value that the $\gamma_2$ equals zero and the whole term in Eq.~(\ref{eq:19}) becomes zero. Thus, the whole Kappa function is equal to zero, which shows that it does not reflect the true behavior of the focused vortex beam. Moreover, the Kappa function encounters a $\pi$-jump in this point. This $z_2$ value can be computed from the formula

\begin{equation}
z_2 = \frac{z_1f_2}{f_2-z_1}
\label{eq:20}
\end{equation}

Fig.~\ref{fig:coeff_A_plots}(a,b) illustrates the problem. As we can see for the $z_2=15.58$ mm (calculated from Eq.~(\ref{eq:20})) the $1/A^{(2)}$ term equals zero.

When the SPP and focusing lens are joined, we only need a coefficient $A^{(1)}$ which has a simple form, free of our problem (at least for the OVSM optical system). This is illustrated in Fig.~\ref{fig:coeff_A_plots}(c,d). The higher order terms $A^{(q)}$ ($q>2$) behave in a similar way. When the SPP and focusing lens work as a single element, they meet the conditions in Eq.~(\ref{eq:9}), for a reasonable OVSM configuration. When $q>2$ the set of conditions in Eq.~(\ref{eq:9}) is more complicated. The detailed study of this problem would explode the volume of this paper, so we will not follow this path. Important thing is that when SPP and focusing lens work separately, condition in Eq.~(\ref{eq:9}) fails for any $q>1$ in case of reasonable OVSM configurations. From practical point of view it is enough to check if the $A^{(q)}$ factor behaves like in Fig.~\ref{fig:coeff_A_plots}(c,d), which is a case in the system shown in Fig.~\ref{fig:two_setups_ab}(b). It is worth noticing here that when using the ISM all $z_{(q)}$ parameters are fixed, so the test for conditions in Eq.~(\ref{eq:9}) is not difficult. When we move any element of our system along the optical axis the conditions in Eq.~(\ref{eq:9}) have to be checked for the whole range of $z_{(q)}$ coordinates.\par
For the reason explained above we cannot use our formulas to the system shown in Fig.~\ref{fig:two_setups_ab}(a). Instead, we have to limit our study to the system shown in Fig.~\ref{fig:two_setups_ab}(b), when both SPP and focusing lens work together. Certainly for the forbidden area the numerical modeling of our optical system is still possible and effective.\par 
We can also find $A^{(q)}$ coefficient inside the first sum in Kappa function. Now, we multiply this terms by $1/A$, located in front of the first sum, so we have

\begin{subequations}
\begin{equation}
\frac{1}{\left(\left(-A^{(q)}\right)^{\frac{3}{2}}\right)^{2n+1}}; \mathrm{for}\, q\, \mathrm{odd}
\label{eq:21a}
\end{equation}
\begin{equation}
\frac{1}{\left(\left(-A^{(q)}\right)^{\frac{3}{2}}\right)^{2n+2}}; \mathrm{for}\, q\, \mathrm{even}
\label{eq:21b}
\end{equation}
\label{eq:21}
\end{subequations}

\begin{figure}
\includegraphics[scale=0.4]{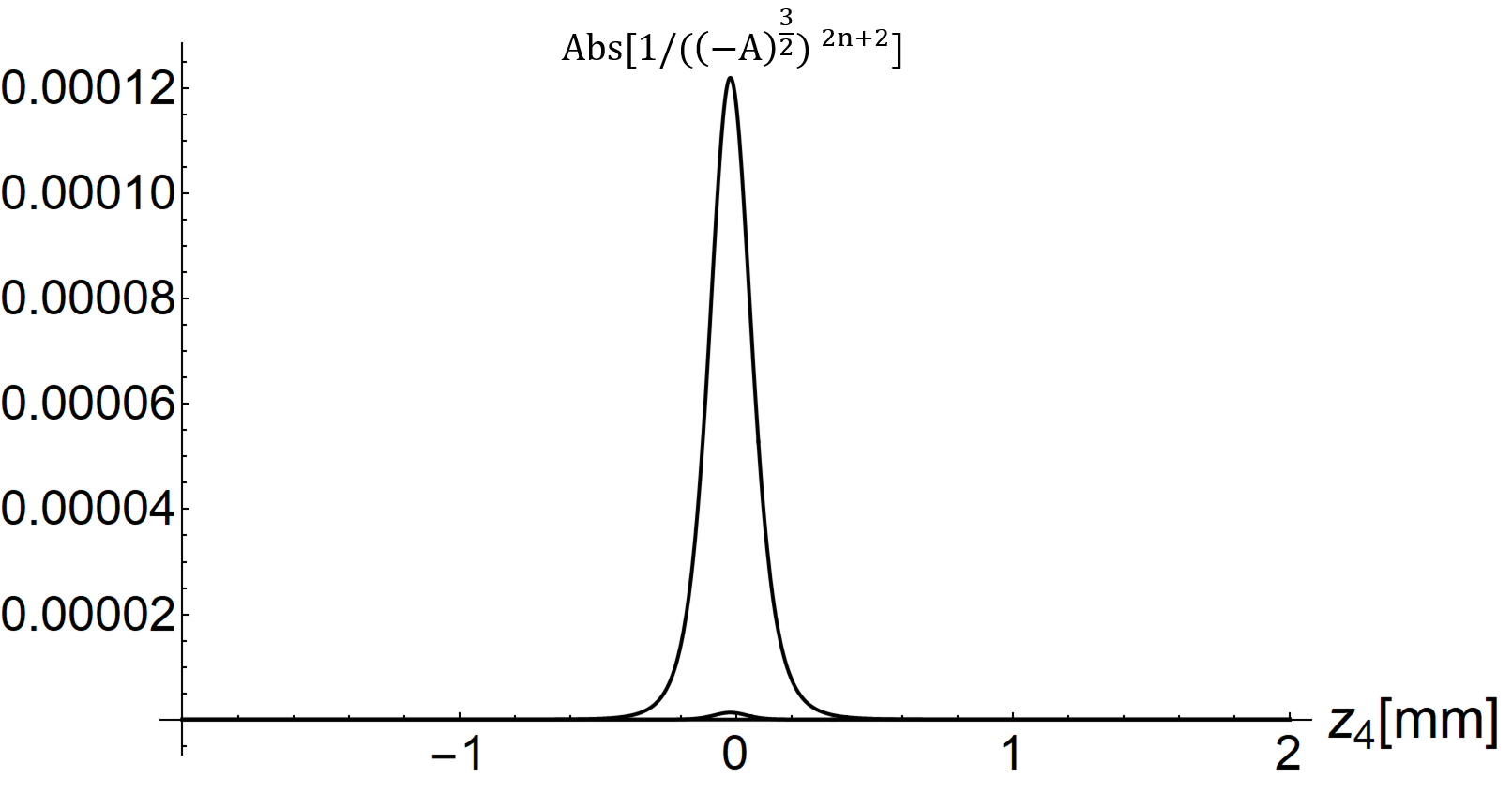}
\caption{The absolute value of the Eq.~(\ref{eq:21b}) containing $A^{(4)}$ coefficient for two values of $n=0$ and $n=1$, in case of even topological charge. For growing $n$ the absolute value drops rapidly. So for $n=1$ the maximum value is $1.2\cdot10^{-4}$ and the plot is hardly visible in this figure. In result the $n=0$ coefficient play a major role in the Kappa function. For $n=2$ the maximum value is $1.3\cdot10^{-6}$. For any next $n$ the maximum value decreases by two orders. \label{fig:abs_A_plot}}
\end{figure}

The first term (for $n=0$) has the largest influence on the phase and amplitude of the Kappa function The next terms rapidly drop in their values. The plot in Fig.~\ref{fig:abs_A_plot} is done for $q=4$, but it represents the typical curve for expressions in Eq.~(\ref{eq:21a})-(\ref{eq:21b}) for any $q$, provided that we avoid the forbidden area defined by conditions (9). As we can see the part for $n=0$ strongly dominates over the part for $n=1$. The next terms (for $n=2,3,...$) are invisible in the figure scale. This domination is particularly strong at the beam center, when $x_{(q)}$ and $y_{(q)}$ coordinates are small (Fig.~\ref{fig:vortex_crossection}). When the $x_{(q)}$ and $y_{(q)}$ become larger, the coefficient grows rapidly with increasing $n$ and things become more complicated. 

\begin{figure}
\includegraphics[scale=0.355]{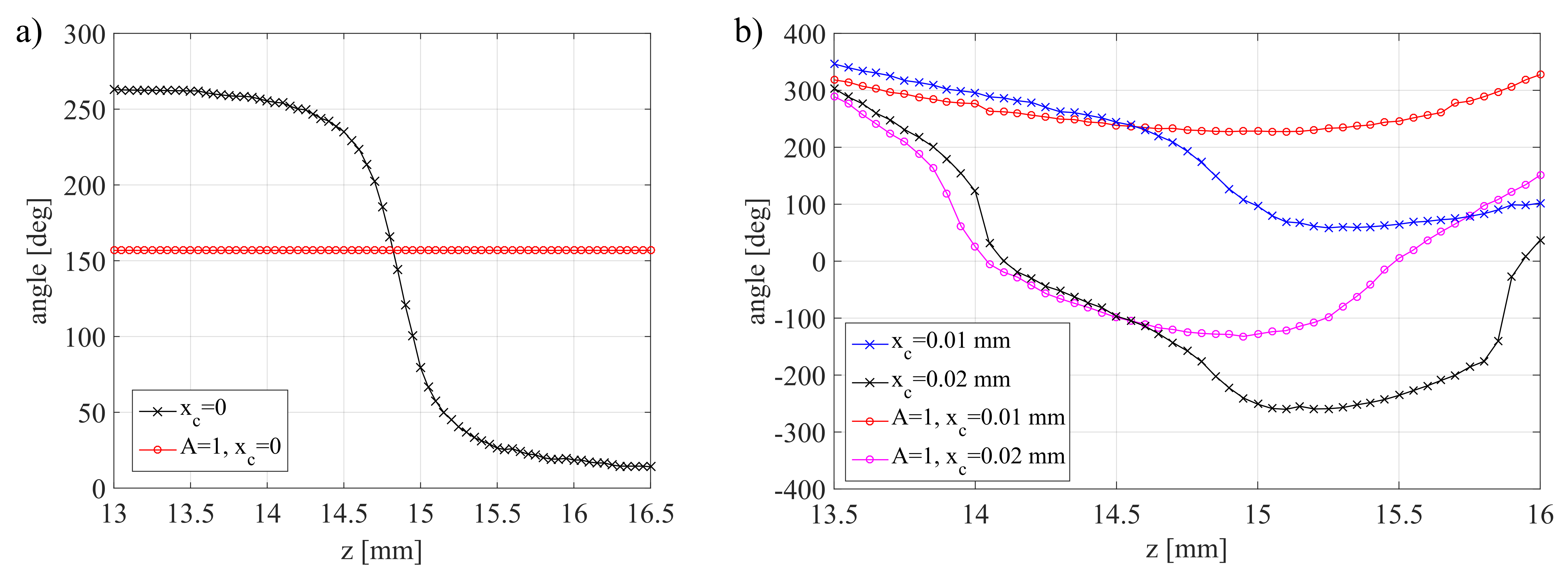}
\caption{The phase rotation in case of $A^{(1)}$ in full form and when $A^{(1)}=1$, but for $x_c=0$; b) The same, but $x_c\neq 0$. \label{fig:phase_rotations}}
\end{figure}

Figure~\ref{fig:coeff_A_plots}(c) suggests that the $A^{(q)}$ coefficient may play a primary role in phase evolution after the SPP, i.e. when changing the position of any element behind the SPP or SPP and focusing lens themselves. This is illustrated in Fig.~\ref{fig:phase_rotations}(a). If $A^{(1)}=1$ and $x_c=0$, there is no phase rotation when changing the position of the focusing lens along the z-axis. We can conclude that in the OVSM system the coefficient $A^{(q)}$ plays, in some respect, the similar role as the Gouy phase in the Gaussian beam. When the $x_c\neq 0$ things become more complicated. The off-axis position of the SPP plate introduces a phase value dependence on $x_c$ being inside the coefficient $B_x$ (Fig.~\ref{fig:phase_rotations}(b)). \par

The other coefficients (but $B_x$, $B_y$) play minor role. The coefficient $B_x$ and $B_y$ are responsible for the vortex trajectory evolution as a function of SPP shift $x_c$. This dependence is linear, but as has been already shown, the direction of vortex trajectory becomes perpendicular to SPP trajectory when the condition in Eq.~(\ref{eq:17}) holds. In the previous papers \cite{50:Plocinniczak,51:Plociniczak} this fact was proved for the small $x_c$. Having the formulas in Eq.~(\ref{eq:11}) we can conclude that they hold for any $x_c$. In paper \cite{53:Popiolek} precise experiments were reported which confirm the theoretical results. \par
The coefficient $\Xi_q$ collects all constant factors. It must be observed when we analyze the rotation of the beam phase while moving the first block in the optical system along the z-axis. When the first element moves away or toward the laser source, the phase of the incident Gaussian beam changes. This incident phase is a part of coefficient $\Xi_q$.\par
\begin{figure}
\includegraphics[scale=0.45]{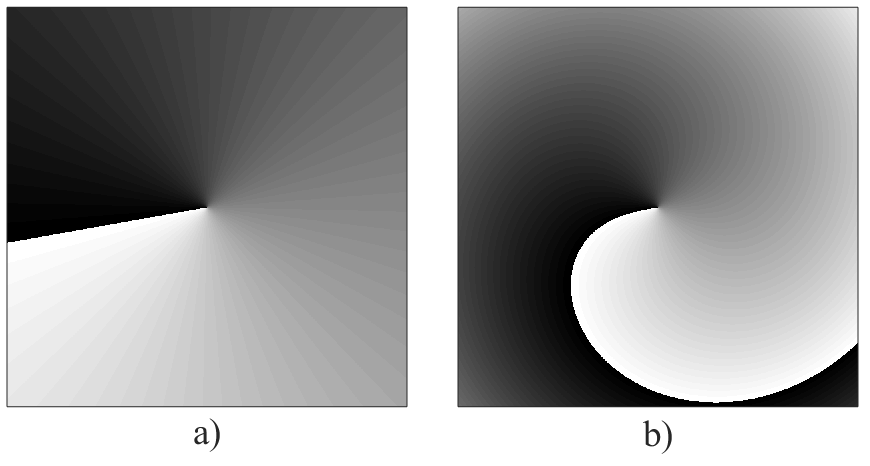}
\caption{The exemplary phase distribution of the vortex beam in case of (a) $C^{(q)}=1$ and (b) full $C^{(q)}$ coefficient. \label{fig:coeff_C}}
\end{figure}

 
The coefficient $C^{(q)}$ multiplies the first sum in function Kappa. Due to the $(x_q^2 + y_q^2)$ factor, the $C$ coefficient is responsible for the equiphase lines curvature (Fig.~\ref{fig:coeff_C}). When $C^{(q)}=1$ the equiphase lines are straight (for $x_c=0$). Since the $C^{(q)}$ coefficient contributes to the Kappa function as $\exp(x_c^2 \alpha + I \Lambda)$, where $\Lambda$ is an imaginary part of the $C$ coefficient, and typically $\alpha \ll \Lambda$ its contribution to the amplitude is small, especially for small $x_c$. 

\section{The ice-skater effect \label{sect:ice-skater_effect}}

To see the effectiveness of our formulas we study the correlation between the focused beam radius and vortex trajectory rotation (see the inset in Fig.~\ref{fig:experimental_setup}). The hypothesis was that the vortex trajectory behaves like a rigid body. Consequently we assumed that in the closed system of a focused vortex beam the angular momentum $L$ is conserved, thus $L = I\omega = const$. Regardless of the assumed model (point mass, flat disk, cylinder, etc.), the moment of inertia $I$ is always proportional to the squared distance $r$ from the rotation axis, i.e. $I \approx r^2$. Therefore, the rotational speed $\omega$ of the vortex trajectory shall satisfy $\omega \approx 1/r^2$. In our case, $\omega$ is the first derivative of the trajectory inclination angle with respect to $z$ ($\omega=d\theta/dz$) and $r$ is the radius of the vortex bright ring, i.e. the distance between points of zero and maximal intensity (e.g. $r=2.4\, \mathrm{\mu}$m for the vortex analyzed in Fig.~\ref{fig:vortex_crossection}). Both the radius of the converging vortex beam and the rotational speed of the trajectory depend on the axial distance $z$. Obviously, as $z$ approaches the focal point, the vortex radius decreases whereas the rotations speed up. As the beam focusses, the radius reaches its minimum and the speed is maximal. This is a direct analogy to the ice skater pulling their arms in for a faster spin. Such a rigid body mechanics approach to the vortex beam has been already studied by Bekshaev et. al. \cite{11:Bekshaev}, for free propagating Gaussian beam with optical vortex. The authors concluded that it is related to the Gouy phase dynamics. We have stated that the coefficient $A$ plays the role of the Gouy phase in our case. Indeed, from Eq.~(\ref{eq:12})-(\ref{eq:13}) we may calculate the angle of the trajectory inclination as

\begin{equation}
\tan\left(\phi_{(q)}\right) = \frac{y_{(q)}}{x_{(q)}} = -2\frac{\alpha}{(-2\beta+k\xi_b^{(q)})}
\label{eq:22}
\end{equation}

For $q=1$ we get

\begin{equation}
\tan\left(\phi_{(q)}\right) = \frac{\alpha}{\beta}
\label{eq:23}
\end{equation}

Calculating the angle of the equiphase line for the $A$ coefficient we get exactly the same formula. \par
\begin{figure}
\includegraphics[scale=0.45]{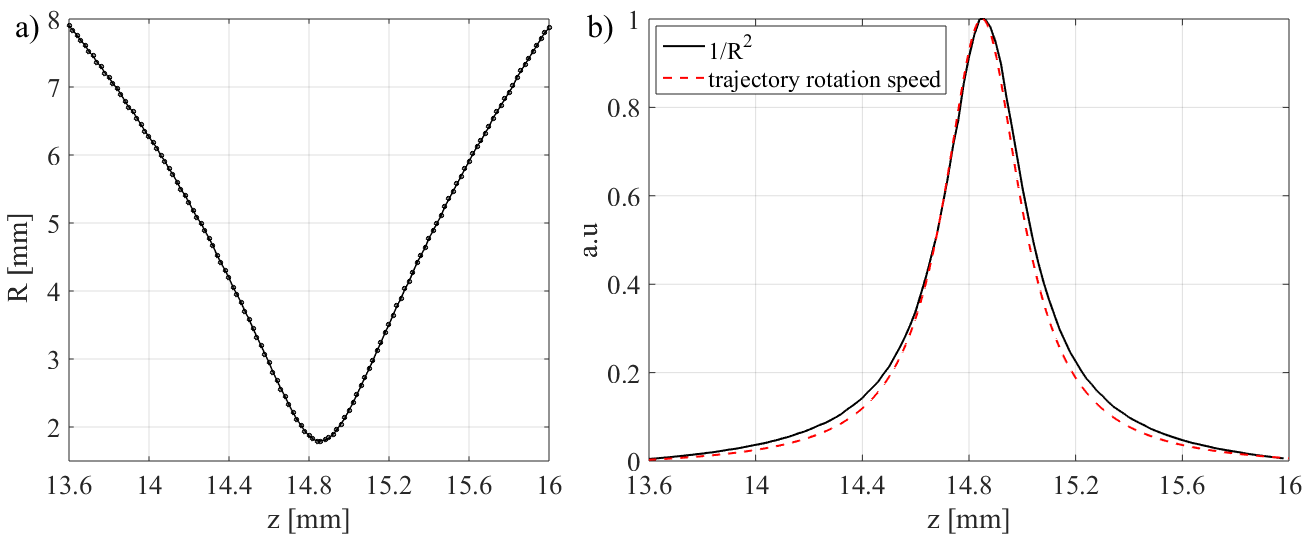}
\caption{a) The radius $R$ of the focused optical vortex ($m=1$) calculated using Kappa function for the setup in Fig.~\ref{fig:two_setups_ab}(b). The radius is minimal at the focal point ($z=14.86$ mm); b) The rotational speed of the vortex trajectory (red) calculated using Eq.~(\ref{eq:24}) compared with the inverse square of $R$ (black). The height of the curves was normalized to $1$. Parameters used in both calculations: $w_0 = 0.4$ mm, $z_G=600$ mm, $\lambda=630$ nm.\label{fig:trajectory_rotations}}
\end{figure}

Here we present the results calculated at the exit of the setup (camera plane) shown in Fig.~\ref{fig:two_setups_ab}(b). The default z-position is where the the focal point of the beam lies very close to the critical plane (plane where the trajectory is perpendicular to the SPP shift $x_c$). Defocusing is performed by moving the focusing objective along the optical axis without changing the position of other elements. Thus, the plane at which the vortex beam is imaged is at different distances from the focus resulting in different vortex radii, as shown in Fig.~\ref{fig:trajectory_rotations}(a). These calculations require running the Kappa function once for each point on the graph. On the other hand, computing the vortex trajectory rotation speed $\omega$ is much easier. In paper \cite{38:Augustyniak} the formula for $\omega$ was given as:

\begin{equation}
\omega = \frac{2kw^2(z)}{4z_0^2+k^2w^4(z)\left(1-z_0\left(\frac{1}{R(z)}+\frac{1}{f}\right)\right)^2}
\label{eq:24}
\end{equation}

In fact the formula in Eq.~(\ref{eq:24}) was derived for the vortex trajectory rotation at the sample plane for the approximated linear case ($n=0$). But the new formulas in Eq.~(\ref{eq:11}) allow to extent their applicability for general case. The classical imaging preserves the vortex trajectory orientation as was shown experimentally in \cite{51:Plociniczak,52:Popiolek}. So we expect that the angle rotation at the image plane is the same as in the sample plane. \par
The obtained relation between the rotational speed and the defocusing is presented in Fig.~\ref{fig:trajectory_rotations}(b), red. Applying the above rigid-body reasoning, $\omega$ was compared with the inverse square of the vortex radius (Fig.~\ref{fig:trajectory_rotations}(b), black). There is a clear agreement between the two curves supporting the hypothesis of rigid-like behavior of vortex trajectory.\par

\section{Conclusions \label{sect:conclusions}}
There is a growing interest in exact theory describing the propagation of the optical vortex beams in optical system, in particular in a system with broken symmetry. The reasons are both understanding the physics of electromagnetic waves, and practical applications. Some examples concerning both science or applications can be found in \cite{11:Bekshaev,14:Bouchal,24:Swartzlander,33:Anzolin,52:Popiolek,60:Ferrando,61:Gibson}, but the list of papers is much longer. In this paper we have enhanced the results presented in our former works \cite{50:Plocinniczak,51:Plociniczak}. We have derived the coefficient for Kappa function for any numbers of elements. We have also enhanced the formulas for the vortex trajectory. Moreover, we have found a closed form of our solution in Eq.~(\ref{eq:11}) using the special function. This new form proved that in classical optical system, the higher order vortices do not split even if the circular symmetry is broken. This is true under the conditions of paraxial approximation. This result is truly surprising. It is hard to check it experimentally. We cannot use an optical system in paraxial approximation. Moreover, any real system introduces errors which break the beam symmetry in a way different than allowed by our theory. This immediately splits the higher order vortices forming a constellation of the first order ones. Nevertheless, the experiment reported in \cite{50:Plocinniczak} suggested the mass center of such a constellation moves as ideal higher order vortex.\par  
The study on $A^{(q)}$ coefficient have shown that it is responsible for breaking the conditions in Eq.~(\ref{eq:9}). In result we cannot analyze the OVSM system with separated SPP plate and focusing lens. The $A^{(q)}$ coefficient has an important role in the rotation of vortex beam phase, which is similar to the role of the Gouy phase in Gaussian beam. The second important factor is the SPP shift $x_c$. When we are close to vortex core we can limit the sum range (in Kappa function) to $n=0$, which is also the result of the $A^{(q)}$ coefficient (Fig.~\ref{fig:abs_A_plot}).\par
We have shown that the vortex trajectory behaves like a rigid body. The range of the trajectory shrinks when the beam converge, so its rotation speed increases according to the rules describing the rigid body behavior. The vortex trajectory rotation is strictly related to the phase rotation of the $A^{(q)}$ coefficient. In \cite{11:Bekshaev} the same relation was shown for free propagation of axial Gaussian vortex beam and the Gouy phase.

\appendix
\section{\label{appendix_A}}
In this section we will prove a recursive formulas in Eq.~(\ref{eq:8}) by mathematical induction.\\
The first step is verification for $q=2$, which has been already done in the previous paper \cite{52:Popiolek}.

\begin{equation}
u_2(x_2,y_2) = \Xi_2 \frac{2i\pi}{k\gamma_2} K(A^{(2)},B_x^{(2)},B_y^{(2)},C^{(2)})
\end{equation}

Next we assume that the formula is correct

\begin{equation}
u_j(x_j,y_j) = \Xi_j \left(\frac{2i\pi}{k}\right)^{j-1} \prod_{s=1}^j \frac{1}{\gamma_s} K(A^{(j)},B_x^{(j)},B_y^{(j)},C^{(j)})
\end{equation}

Finally, we have to prove the implication

\begin{equation}
u_j(x_j,y_j) \Rightarrow u_{j+1}(x_{j+1},y_{j+1})
\end{equation}

We can write a right side of the implication as

\begin{equation}
\begin{split}
u_{j+1}(x_{j+1},y_{j+1}) & = \xi_{j+1} \iint \displaylimits_{\mathbb{R}^2} u_j(x_j,y_j)\, e^{\frac{ik}{2z_{j+1}}(x_{j+1}^2+y_{j+1}^2)}\, e^{\frac{ik}{2z_{j+1}}(x_j^2+y_j^2)}\, e^{-\frac{ik}{2f_{j+1}}(x_j^2+y_j^2)}\, \\ & \times e^{-\frac{ik}{z_{j+1}}(x_{j+1}x_j+y_{j+1}y_j)} \mathrm{d}x_j \mathrm{d}y_j
\end{split}
\end{equation}

After plugging the definition of $u_j$ into the integral to obtain

\begin{equation}
\begin{split}
u_{j+1}(x_{j+1},y_{j+1}) = \xi_{j+1} \iint \displaylimits_{\mathbb{R}^2} \Xi_j \left(\frac{2i\pi}{k}\right)^{j+1} \prod_{s=1}^j \frac{1}{\gamma_s} \Bigg[\iint \displaylimits_{\mathbb{R}^2} e^{imArg(x_0+iy_0)}\, \\ \times e^{-\left(\frac{1}{w^2(z)}+\frac{ik}{2R(z)}\right) [(x_0-x_c)^2+y_0^2]}\,  e^{-\frac{ik}{2f_1}[(x_0-x_c)^2+y_0^2]}\, e^{\frac{ik}{2z_1}[(x_0-x_c)^2+y_0^2]} \\ \times e^{-\frac{ik}{2\gamma_2}\left[\left(\frac{x_0-x_c}{z_1}\right)^2+\left(\frac{y_0}{z_1}\right)^2\right]}\,  e^{-\frac{ik}{2\gamma_3}\left[\left(\frac{x_0-x_c}{z_1z_2\gamma_2}\right)^2+\left(\frac{y_0}{z_1z_2\gamma_2}\right)^2\right]}\, e^{-\frac{ik}{2\gamma_4}\left[\left(\frac{x_0-x_c}{z_1z_2z_3\gamma_2\gamma_3}+\frac{x_4}{z_4}\right)^2+\left(\frac{y_0}{z_1z_2z_3\gamma_2\gamma_3}+\frac{y_4}{z_4}\right)^2\right]} \\ \times \dots e^{-\frac{ik}{2\gamma_{j-1}}\left[\left(\frac{x_0-x_c}{\prod_{s=1}^{j-2}z_s\gamma_s}\right)^2+\left(\frac{y_0}{\prod_{s=1}^{j-2}z_s\gamma_s}\right)^2\right]}\, e^{-\frac{ik}{2\gamma_j}\left[\left(\frac{x_0-x_c}{\prod_{s=1}^{j-1}z_s\gamma_s}+\frac{x_j}{z_j}\right)^2+\left(\frac{y_0}{\prod_{s=1}^{j-1}z_s\gamma_s}+\frac{y_j}{z_j}\right)^2\right]}\, \mathrm{d}x_0 \mathrm{d}y_0\Bigg] \\ \times e^{\frac{ik}{2z_{j+1}}(x_{j+1}^2+y_{j+1}^2)}\, e^{\frac{ik}{2z_{j+1}}(x_j^2+y_j^2)}\, e^{-\frac{ik}{2f_{j+1}}(x_j^2+y_j^2)}\, e^{-\frac{ik}{z_{j+1}}(x_{j+1}x_j+y_{j+1}y_j)} \mathrm{d}x_j \mathrm{d}y_j
\end{split}
\end{equation}

By interchanging the order of integration we can write

\begin{equation}
\begin{split}
u_{j+1}(x_{j+1},y_{j+1}) = \Xi_{j+1} \left(\frac{2i\pi}{k}\right)^{j+1} \prod_{s=1}^j \frac{1}{\gamma_s} \iint \displaylimits_{\mathbb{R}^2}  e^{imArg(x_0+iy_0)}\, \\ \times e^{-\left(\frac{1}{w^2(z)}+\frac{ik}{2R(z)}\right) \cdot [(x_0-x_c)^2+y_0^2]}\, e^{-\frac{ik}{2f_1}[(x_0-x_c)^2+y_0^2]}\, e^{\frac{ik}{2z_1}[(x_0-x_c)^2+y_0^2]}\, e^{\frac{ik}{2z_{j+1}}(x_{j+1}^2+y_{j+1}^2)}\, \\ \times e^{-\frac{ik}{2\gamma_2}\left[\left(\frac{x_0-x_c}{z_1}\right)^2+\left(\frac{y_0}{z_1}\right)^2\right]}\, e^{-\frac{ik}{2\gamma_3}\left[\left(\frac{x_0-x_c}{z_1z_2\gamma_2}\right)^2+\left(\frac{y_0}{z_1z_2\gamma_2}\right)^2\right]}\, e^{-\frac{ik}{2\gamma_4}\left[\left(\frac{x_0-x_c}{z_1z_2z_3\gamma_2\gamma_3}\right)^2+\left(\frac{y_0}{z_1z_2z_3\gamma_2\gamma_3}\right)^2\right]} \\ \times \dots e^{-\frac{ik}{2\gamma_{j-1}}\left[\left(\frac{x_0-x_c}{\prod_{s=1}^{j-2}z_s\gamma_s}\right)^2+\left(\frac{y_0}{\prod_{s=1}^{j-2}z_s\gamma_s}\right)^2\right]}\, e^{-\frac{ik}{2\gamma_j}\left[\left(\frac{x_0-x_c}{\prod_{s=1}^{j-1}z_s\gamma_s}\right)^2+\left(\frac{y_0}{\prod_{s=1}^{j-1}z_s\gamma_s}\right)^2\right]}\, \\ \times \Bigg[\iint \displaylimits_{\mathbb{R}^2} e^{-\frac{ik}{2\gamma_jz_j}\left(\frac{2(x_0-x_c)x_j}{\prod_{s=1}^{j-1}z_s\gamma_s}+\frac{2y_0y_j}{\prod_{s=1}^{j-1}z_s\gamma_s}\right)}\, e^{-\frac{ik}{2z_j^2\gamma_j} (x_j^2+y_j^2)}\, \\ \times  e^{\frac{ik}{2z_{j+1}}(x_j^2+y_j^2)}\, e^{-\frac{ik}{2f_{j+1}}(x_j^2+y_j^2)}\, e^{-\frac{ik}{z_{j+1}}(x_{j+1}x_j+y_{j+1}y_j)} \mathrm{d}x_j \mathrm{d}y_j\Bigg] \mathrm{d}x_0 \mathrm{d}y_0
\end{split}
\end{equation}

The integral in the parentheses is Gaussian and thus can be computed explicitly using the formula 

\begin{equation}
\int_{-\infty}^{\infty} e^{i\sigma x^2-i\mu x } \mathrm{d}x = \sqrt{\frac{\pi}{-i\sigma}} e^{-\frac{i\mu^2}{4\sigma}}
\end{equation}

Eventually, we can write

\begin{equation}
u_{j+1}(x_{j+1},y_{j+1}) = \Xi_{j+1} \left(\frac{2i\pi}{k}\right)^j \prod_{s=1}^{j+1} \frac{1}{\gamma_s} K(A^{(j+1)},B_x^{(j+1)},B_y^{(j+1)},C^{(j+1)})
\end{equation}

\section{\label{appendix_B}}
In this section we will prove the formula in Eq.~(\ref{eq:10}).\\
In the first step the formula for positive vortex charge can be written as

\begin{equation}
u_{j+}(x_j,y_j) = \Xi_j K(A^{(j)},B_x^{(j)},B_y^{(j)},C^{(j)})
\end{equation}

Similarly, the formula for negative vortex charge is

\begin{equation}
u_{j-}(x_{j},y_{j}) = \Xi_j \iint \displaylimits_{R^2} e^{im\phi}\, e^{A^{(j)}\rho^2+B_x^{(j)}\rho\cos\phi+B_y^{(j)}\rho\sin\phi+C^{(j)}}\, \rho \mathrm{d}\rho \mathrm{d}\phi
\end{equation}

Then, we substitute $\theta=2\pi-\phi$ for integration over $\theta$, we have

\begin{equation}
\Xi_j\, e^{im2\pi} \iint \displaylimits_{R^2} e^{-im\theta}\, e^{A^{(j)}\rho^2+B_x^{(j)}\rho\cos\theta-B_y^{(j)}\rho\sin\theta+C^{(j)}}\, \rho \mathrm{d}\rho \mathrm{d}\theta
\end{equation}

Finally, we can write

\begin{equation}
u_{j-}(x_j,y_j) = -e^{-im2\pi}\, \Xi_j\, K(A^{(j)},B_x^{(j)},-B_y^{(j)},C^{(j)})
\end{equation}

\section{\label{appendix_C}}
\
In this section we will prove the formula in Eq.~(\ref{eq:11}).\\
Using polar coordinates in formula Eq.~(\ref{eq:1}) we have

\begin{equation}
u(x_1,y_1) = \int_{0}^{2\pi} e^{im\phi} \mathrm{d}\phi \int_{0}^{\infty} e^{A\rho^2+B_x\rho\cos\phi+B_y\rho\sin\phi+C}\, \rho \mathrm{d}\rho
\end{equation}

Integrating over the radius we have

\begin{equation}
\int_{0}^{\infty} e^{-\gamma \cdot \rho^2-\sigma\rho}\, \rho \mathrm{d}\rho = \frac{1}{2\gamma} \left[1-\sqrt{\pi}\frac{\sigma}{2\sqrt{\gamma}}\, e^{\frac{\sigma^2}{4\gamma}} \left(1-erf\left(\frac{\sigma}{2\sqrt{\gamma}}\right)\right)\right]
\end{equation}

After integration we get

\begin{equation}
\begin{split}
u(x_1,y_1) & = -\frac{e^C}{2A} \int_{0}^{2\pi} \bigg[1-\sqrt{\pi}\frac{-(B_x\cos\phi+B_y\sin\phi)}{2\sqrt{-A}}\, e^{-\frac{(B_x\cos\phi+B_y\sin\phi)^2}{4A}} \\ & \times \left(1-erf\left(\frac{-(B_x\cos\phi+B_y\sin\phi)}{2\sqrt{-A}}\right)\right)\bigg]\, e^{im\phi}\, \mathrm{d}\phi
\end{split}
\end{equation}

After calculating the sum we have

\begin{equation}
\begin{split}
u(x_1,y_1) & = -\frac{\sqrt{\pi}e^{C}}{2\sqrt{-A}} \sum_{n=\frac{m-1}{2}}^\infty \frac{1}{n!}\, \frac{\pi}{2^{2n}} {{2n+1}\choose{n+\frac{m+1}{2}}} \\ & \times \left(\frac{B_x}{2\sqrt{-A}}+i\frac{B_y}{2\sqrt{-A}}\right)^{n+\frac{m+1}{2}}  \left(\frac{B_x}{2\sqrt{-A}}-i\frac{B_y}{2\sqrt{-A}}\right)^{n-\frac{m-1}{2}}; \mathrm{for}\, m\, \mathrm{odd}
\end{split}
\end{equation}

\begin{equation}
\begin{split}
u(x_1,y_1) & = \frac{e^{C}}{2\sqrt{-A}} \sum_{n=\frac{m}{2}-1}^\infty \frac{2^{n+1}}{(2n+1)!!}\, {{2n+2}\choose{n+1+\frac{m}{2}}} \\ & \times \left(\frac{B_x}{2\sqrt{-A}}+i\frac{B_y}{2\sqrt{-A}}\right)^{n+1+\frac{m}{2}}  \left(\frac{B_x}{2\sqrt{-A}}-i\frac{B_y}{2\sqrt{-A}}\right)^{n+1-\frac{m}{2}}; \mathrm{for}\, m\, \mathrm{even}
\end{split}
\end{equation}

Finally, we can write

\begin{equation}
\begin{split}
u(x_1,y_1) & = -\frac{\pi\sqrt{\pi}e^{C}}{2^m\sqrt{-A}} \frac{1}{\left(\frac{m-1}{2}\right)!}\, \left(\frac{B_x}{2\sqrt{-A}}+i\frac{B_y}{2\sqrt{-A}}\right)^m \\ & \times _1F_1\left(1+\frac{m}{2},1+m,\left(\frac{B_x}{2\sqrt{-A}}\right)^2+\left(\frac{B_y}{2\sqrt{-A}}\right)^2\right); \mathrm{for}\, m\, \mathrm{odd}
\end{split}
\end{equation}

\begin{equation}
\begin{split}
u(x_1,y_1) & = \frac{\pi e^{C}}{2^{\frac{m}{2}}\sqrt{-A}} \frac{1}{(m-1)!!}\, \left(\frac{B_x}{2\sqrt{-A}}+i\frac{B_y}{2\sqrt{-A}}\right)^m \\ & \times _1F_1\left(1+\frac{m}{2},1+m,\left(\frac{B_x}{2\sqrt{-A}}\right)^2+\left(\frac{B_y}{2\sqrt{-A}}\right)^2\right); \mathrm{for}\, m\, \mathrm{even}
\end{split}
\end{equation}

\bibliographystyle{apsrev4-2}
\bibliography{bibliography}

\end{document}